# Robust Blind Source Separation by Soft Decision-Directed Non-Unitary Joint Diagonalization

Wenjuan Liu [a], Dazheng Feng [b,*], Bingnan Pei [b], Mengdao Xing[b], Xinhong Meng[b], Qianru Wei[c]

[a] Xidian University

[b] Fujian Key Laboratory of the Modern Communication and Beidou Positioning Technology in Universities, Quanzhou University of Information Engineering

[c] School of Software and Microelectronics, Northwestern Polytechnical University, Xi'an, China.

**Abstract:** Approximate joint diagonalization of a set of matrices provides a powerful framework for numerous statistical signal processing applications. For non-unitary joint diagonalization (NUJD) based on the least-squares (LS) criterion, outliers, also referred to as anomaly or discordant observations, have a negative influence on the performance, since squaring the residuals magnifies the effects of them. To solve this problem, we propose a novel cost function that incorporates the soft decision-directed scheme into the least-squares algorithm and develops an efficient algorithm. The influence of the outliers is mitigated by applying decision-directed weights which are associated with the residual error at each iterative step. Specifically, the mixing matrix is estimated by a modified stationary point method, in which the updating direction is determined based on the linear approximation to the gradient function. Simulation results demonstrate that the proposed algorithm outperforms conventional non-unitary diagonalization algorithms in terms of both convergence performance and robustness to outliers.

## 1. Introduction

Approximate joint diagonalization (AJD) of a set of square matrices is a member of the most dynamic topic in signal processing, which is found various applications in wireless communication, biomedical signal processing and speech processing, see [1] and the references therein. Mathematically, AJD can be stated as follows: Given a set of matrices $\{\mathbf{R}_1, \mathbf{R}_2, \ldots, \mathbf{R}_K\}$, the AJD problem consists of seeking a nonsingular mixing matrix $\mathbf{A} \in \mathbb{C}^{M \times N}$ and the associated diagonal matrices $\mathbf{D}_k \in \mathbb{C}^{N \times N}$, $k = 1, 2, \ldots, K$ such that the following common structures are best fitted:

$$\mathbf{R}_k = \mathbf{A}\mathbf{D}_k\mathbf{A}^{\mathrm{H}}, k = 1, 2, \ldots, K, \quad (1)$$

where superscript $(\cdot)^{\mathrm{H}}$ denotes Hermitian transpose. Or AJD searches the corresponding diagonalizing matrix $\mathbf{V}(= \mathbf{A}^{-\dagger})$ such that the transformed matrices

$$\mathbf{D}_k = \mathbf{V}\mathbf{R}_k\mathbf{V}^{\mathrm{H}}, k = 1, 2, \ldots, K \quad (2)$$

are as diagonal as possible. Here $(\cdot)^{-\dagger}$ stands for the pseudo-inverse of the matrix argument or directly the inverse in the square case.

*Corresponding author.
E-mail address: dzfeng@xidian.edu.cn (D.-Z. Feng).

In the context of blind source separation (BSS), various algorithms have been developed to solve the AJD problem. To name a few, the algorithms [2]–[8] search a non-singular matrix $\mathbf{V}$ that minimizes the following contrast criterion:

$$\min J_{\mathrm{off}}(\mathbf{V}) = \sum_{k=1}^{K} \omega_k \sum_{m=1}^{N} \sum_{n=1, n \neq m}^{N} \left| \left[\mathbf{V}\mathbf{R}_k\mathbf{V}^{\mathrm{H}}\right]_{mn} \right|^2 \quad (3)$$

where $[\mathbf{R}]_{mn}$ denotes the entry with row index $m$ and column index $n$ in a matrix $\mathbf{R}$. Eq. (3) is obviously minimized by $\mathbf{V} = \mathbf{0}$. Therefore, it must resort to an appropriate constraint or a penalty term to avoid the trivial solution in the non-unitary case. The algorithms proposed in [2], [3] constrains the updating matrix in each iterative step to be strictly diagonal dominant, to make sure the diagonalizing matrix is invertible. And in [6] the FAJD algorithm adopts a penalty term which is proportional to $\log|\det(\mathbf{V})|$ to avoid the trivial and degenerate solution. The algorithms proposed in [9]–[12] build a least-squares criterion

$$\min J_{\mathrm{LS}}\left(\mathbf{A}, \{\mathbf{D}_k\}_{k=1}^{K}\right) = \sum_{k=1}^{K} \omega_k \left\| \mathbf{R}_k - \mathbf{A}\mathbf{D}_k\mathbf{A}^{\mathrm{H}} \right\|_{\mathrm{F}}^2 \quad (4)$$

which search for not only the mixing matrix $\mathbf{A}$ but also a set of diagonal matrices $\mathbf{D}_k, k = 1, 2, \ldots, K$. In [11], considering $\mathbf{A}$ and $\mathbf{A}^{\mathrm{H}}$ as different variables, the LS criterion (4) can be considered as a quadratic function with respect to $\mathbf{A}$, $\mathbf{A}^{\mathrm{H}}$ and $\mathbf{D}_k, k = 1, 2, \ldots, K$, respectively, which could be minimized by an alternating least squares (ALS) procedure. The maximum likelihood (ML) approach developed by Pham [13] minimizes an approximate log-likelihood

criterion

$$\min J_{LL}(\mathbf{V}) = \sum_{k=1}^{K} \log \frac{\det\left(\text{diag}\left\{\mathbf{V}\mathbf{R}_k\mathbf{V}^H\right\}\right)}{\det\left(\mathbf{V}\mathbf{R}_k\mathbf{V}^H\right)} \quad (5)$$

where $\text{diag}\{\cdot\}$ sets the off-diagonal entries of the matrix argument to zero. Criterion (5) is scale-invariant in $\mathbf{V}$ and does not require any constraints. However, it is meaningful only for positive definite target-matrices.

The existing algorithms for AJD generally assume that the target matrices admit a jointly diagonalizable structure. However, the situation where there exist outliers in the target matrices set has never been examined. Outliers here refer to matrices that contain random errors with extreme values or do not follow the diagonalizable structure. It is known that detection and prediction of outliers are challenging tasks [14]. Unfortunately, the LS-based algorithms are sensitive to outliers [15]. Thus, it is still of interest to search for a more robust method to reduce the influence of the outliers and improve the performance of NUJD even further. To the best of our knowledge, this issue has not been discussed in the literature. The main purpose of this paper is to fill this important gap.

To address the above challenge, we proposed a novel soft decision-directed algorithm for non-unitary joint diagonalization. When compared with the existing NUJD algorithms, the main advantages of the presented algorithm are as follows: (i) we introduce an appropriate cost function for integrating the least-squares algorithm with the SDD scheme; (ii) it can directly apply in the rectangular mixing matrix without any dimension reduction procedure; (iii) it makes use of a modified stationary point method for the estimation of the mixing matrix based on a linear approximation to the gradient function of the criterion, which is efficient and simple. Computer simulations are provided to illustrate that the proposed algorithm is robust and computationally efficient.

## 2. Soft Decision-Directed Method

It is well known that the least-squares method works best for data that does not contain a large number of random fitting errors although it is not assumed normally distributed errors [16]. If it doesn't match this assumption, the fit may be unduly influenced by data of poor quality. To improve the performance of least-squares method, one can use weighted least-squares (WLS) regression where additional scale factors, i.e. the weights $\omega_k$ in Eqs. (3) and (4), should be included in the process. The weights provided in the fitting procedure correctly indicate the different levels of quality presented in the observations and adjust the amount of influence each data point has on the final parameter estimation. The weights are difficult to achieve both in theoretical derivation and practical applications. In this paper, we propose a novel cost function which incorporates the soft decision-directed scheme [17], [18] into the least-squares method. The resulted soft decision-directed joint diagonalization (SDDJD) algorithm estimates the mixing matrix $\mathbf{A}$ and diagonal matrices by maximizing the following criterion:

$$J\left(\mathbf{A},\{\mathbf{D}_k\}_{k=1}^K\right) = \log\left[\sum_{k=1}^K \exp\left(-\frac{\|\mathbf{R}_k - \mathbf{A}\mathbf{D}_k\mathbf{A}^H\|_F^2}{2\sigma^2}\right)\right] \quad (6)$$

where $\sigma^2$ is decided by the residual error $\|\mathbf{R}_k - \mathbf{A}\mathbf{D}_k\mathbf{A}^H\|_F^2$. The motivation for taking this cost function is as follows: we would like to be able to quantify the fitting error and subsequently make precise revisions of $\sigma^2$.

Roughly speaking, the proposed algorithm is built on a two-stage iterative scheme. The first stage is to estimate the mixing matrix $\mathbf{A}$ when the diagonal matrices $\mathbf{D}_k, k=1,2,\ldots,K$ are fixed; while the second stage is to estimate the diagonal matrices $\mathbf{D}_k, k=1,2,\ldots,K$ when $\mathbf{A}$ is fixed.

### 2.1. Updating of mixing matrix

In the first stage, an optimal solution is not easily found but can be achieved by an optimization algorithm. Considering the relation $\operatorname{tr}(\mathbf{U}\mathbf{U}^H) = \operatorname{tr}(\mathbf{U}^H\mathbf{U})$, the conjugate gradient of Eq. (6) with respect to $\mathbf{A}$ is

$$\frac{\partial J(\mathbf{A})}{\partial \mathbf{A}^*} = \sum_{k=1}^K \mu_k \sigma^{-2} \mathbf{A}\left(\mathbf{D}_k^* \mathbf{A}^H \mathbf{A} \mathbf{D}_k + \mathbf{D}_k \mathbf{A}^H \mathbf{A} \mathbf{D}_k^*\right) \\ - \sum_{k=1}^K \mu_k \sigma^{-2} \left(\mathbf{R}_k^H \mathbf{A} \mathbf{D}_k + \mathbf{R}_k \mathbf{A} \mathbf{D}_k^*\right) \quad (7)$$

where

$$\mu_k = \frac{\exp\left(-\|\mathbf{R}_k - \mathbf{A}\mathbf{D}_k\mathbf{A}^H\|_F^2 / 2\sigma^2\right)}{\sum_{l=1}^K \exp\left(-\|\mathbf{R}_l - \mathbf{A}\mathbf{D}_l\mathbf{A}^H\|_F^2 / 2\sigma^2\right)}. \quad (8)$$

If target matrix $\mathbf{R}_k$ has a higher residual error $\|\mathbf{R}_k - \mathbf{A}\mathbf{D}_k\mathbf{A}^H\|_F^2$ than $\mathbf{R}_l$, then the weight $\mu_k$ is lower than $\mu_l$. In other words, the more reliable value may have a larger influence on the gradient direction. $\mu_k$ in (8) is useful in very many practical situations, since there is no chance to get knowledge about outliers or the weight to each of the recorded data point directly. Notice that expression (8) can be regarded as the weight to each term of the gradient (3.9) in BIA [10]. Similarly to BIA method, we firstly search a linear approximation to gradient function (7) near the previous value $\mathbf{A}_{(t-1)}$. Secondly, letting this approximation function be equal to zero yields a set of linear equations

$$\nabla_{\mathbf{A}_{(t)}} J(\mathbf{A}_{(t)}) \approx \mathbf{A}_{(t)} \mathbf{C}(\mathbf{A}_{(t-1)}) - \mathbf{B}(\mathbf{A}_{(t-1)}) = \mathbf{0}. \tag{9}$$

where

$$\mathbf{C}(\mathbf{A}) = \sum_{k=1}^{K} \mu_k \left( \mathbf{D}_k^* \mathbf{A}^H \mathbf{A} \mathbf{D}_k + \mathbf{D}_k \mathbf{A}^H \mathbf{A} \mathbf{D}_k^* \right), \tag{10}$$

$$\mathbf{B}(\mathbf{A}) = \sum_{k=1}^{K} \mu_k \left( \mathbf{R}_k^H \mathbf{A} \mathbf{D}_k + \mathbf{R}_k \mathbf{A} \mathbf{D}_k^* \right). \tag{11}$$

And the current $\mathbf{A}_{(t)}$ is thirdly obtained by solving such a set of linear equations (9) as follows:

$$\mathbf{A}_{(t)} = \mathbf{B}(\mathbf{A}_{(t-1)}) \mathbf{C}^{-1}(\mathbf{A}_{(t-1)}) \tag{12}$$

***Remark***: The parameter $\sigma^2$ significantly affects the performance, detailed discussion on the influence of $\sigma^2$ on the performance of the fitting can be found in [19], [20]. In this paper, we set $\sigma^2 = \sum_{k=1}^{K} \mu_k \|\mathbf{R}_k - \mathbf{A} \mathbf{D}_k \mathbf{A}^H\|_F^2$. It can be interpreted as a weighted average of residual errors at each iteration step.

### 2.2. Updating of diagonal matrices

In the second stage, the diagonal matrices are determined as following: Firstly, given matrix $\mathbf{A}$, Eq.(6) can be expressed as:

$$J\left(\{\mathbf{D}_k\}_{k=1}^{K}\right) = \log \left[ \sum_{k=1}^{K} \exp\left( -\frac{\|\mathbf{r}_k - (\mathbf{A}^* \oslash \mathbf{A}) \mathbf{d}_k\|_F^2}{2\sigma^2} \right) \right] \tag{13}$$

where $\mathbf{r}_k = \text{vec}(\mathbf{R}_k)$, $\oslash$ denotes the Kathri-Rao (column-wise Kronecker) product, i.e.,

$\mathbf{U} \oslash \mathbf{V} = [\mathbf{u}_1 \otimes \mathbf{v}_1, \mathbf{u}_2 \otimes \mathbf{v}_2, \ldots, \mathbf{u}_N \otimes \mathbf{v}_N] \in \mathbb{C}^{M^2 \times N}$, $\mathbf{d}_k = \text{diag}(\mathbf{D}_k) = [\lambda_{k,1}, \lambda_{k,2}, \ldots, \lambda_{k,N}]^T$ and $\lambda_{k,n}$ are the element $(n, n)$ of the diagonal matrix $\mathbf{D}_k$. Let the conjugate gradient of Eq. (13) with respect to $\mathbf{d}_k$, $k = 1, 2, \ldots, K$ be equal to zero, we have

$$\frac{\partial J\left(\{\mathbf{D}_k\}_{k=1}^K\right)}{\partial \mathbf{d}_k^*} = \mu_k \frac{\left(\mathbf{A}^* \oslash \mathbf{A}\right)^H \left[\mathbf{r}_k - \left(\mathbf{A}^* \oslash \mathbf{A}\right)\mathbf{d}_k\right]}{\sigma^2} = \mathbf{0} \tag{14}$$

which is equivalent to $\left(\mathbf{A}^* \oslash \mathbf{A}\right)^H \left[\mathbf{r}_k - \left(\mathbf{A}^* \oslash \mathbf{A}\right)\mathbf{d}_k\right] = \mathbf{0}$. Eq. (14) then could be given into a compact form:

$$\mathbf{H}\mathbf{d}_k = \mathbf{b}_k, k = 1, 2, \ldots, K \tag{15}$$

where $\mathbf{H} = (\mathbf{A}^H \mathbf{A}) \odot (\mathbf{A}^H \mathbf{A})^* \in \mathbb{C}^{N \times N}$, $\odot$ denotes the Hadamard product and $\mathbf{b}_k = [\mathbf{a}_1^H \mathbf{R}_k \mathbf{a}_1, \mathbf{a}_2^H \mathbf{R}_k \mathbf{a}_2, \ldots, \mathbf{a}_N^H \mathbf{R}_k \mathbf{a}_N]^T \in \mathbb{C}^{N \times 1}$. One can easily find from Eq. (15) that the estimation of $\mathbf{D}_k$ could be written as:

$$\mathbf{D}_k = \text{diag}\left(\mathbf{H}^{-1}\mathbf{b}_k\right), k = 1, 2, \ldots, K \tag{16}$$

where $\text{diag}(\cdot)$ denotes the operator, which builds a diagonal matrix from its argument.

The proposed algorithm alternates between the aforementioned two stages until an acceptable maximization is attained. Note that, in the implementation, the diagonal matrices could be updated after a few iterations given in Eq. (12) until the given maximum number of iterations is reached, or the change of the mixing matrix is smaller than a predefined threshold.

## 3. Simulation results

This section is devoted to the performance evaluation of the proposed SDDJD algorithm in comparison with six state-of-the-art algorithms namely, BIA [10], FAJD[6], NOODLES [3], ALS [11], CJDi [4] and UWEDGE [12]. To measure the performance of each algorithm, we use three classical performance indexes: the computation time (CPU time) and the number of iterations required to reach convergence, as well as global rejection level (GRL) [3], [4], [10] which can be expressed as:

$$\text{GRL} = \sum_{n=1}^{N}\left(\frac{\sum_{m=1}^{N}|G_{nm}|^2}{\max_p|G_{np}|^2}-1\right)+\sum_{n=1}^{N}\left(\frac{\sum_{m=1}^{N}|G_{mn}|^2}{\max_p|G_{pn}|^2}-1\right) \quad (17)$$

where $\mathbf{G}=\hat{\mathbf{A}}^{-\dagger}\mathbf{A}$ or $\mathbf{G}=\hat{\mathbf{V}}\mathbf{A}$ whose $(m,n)$-th element is $G_{mn}$, $\hat{\mathbf{A}}$ is the mixing matrix estimated by ALS, BIA and SDDJD, $\hat{\mathbf{V}}$ is the demixing matrix obtained by FAJD, CJDi, UWEDGE and NOODLES. The comparing algorithms are implemented in MATLAB R2016b environment on a PC configured with Intel(R) Core (TM) i5-6300HQ CPU @ 2.30GHz, 12GB memory. In all experiments, if the condition $|\hat{\mathbf{A}}_{(t)}-\hat{\mathbf{A}}_{(t-1)}|<\varepsilon$ ($|\hat{\mathbf{V}}_{(t)}-\hat{\mathbf{V}}_{(t-1)}|<\varepsilon$) is satisfied, where $\hat{\mathbf{A}}_{(t)}$ and $\hat{\mathbf{V}}_{(t)}$ denotes the estimated mixing and unmixing matrices associated with the $t$-th iterative step respectively, or $t>T_{\max}$, these algorithms are considered to have converged and will be stopped. The present toleration $\varepsilon$ is taken as $10^{-4}$ and the maximum iterative number $T_{\max}$ is 1500. All the results below are calculated over 100 independent trials.

In the following experiments, a set of $M\times M$ joint diagonalizable matrices are given by $\mathbf{R}_k=\mathbf{A}\mathbf{D}_k\mathbf{A}^H+\delta\Delta\mathbf{R}_k$ where $\delta$ is a positive number tuning the noise level measured by $\text{NER}=10\log_{10}(1/\delta^2)$, the outliers are given as $\mathbf{R}_k=\Delta\mathbf{R}_k$. The $M\times N$ non-unitary mixing matrix $\mathbf{A}$, each of error matrices $\Delta\mathbf{R}_k$ and diagonal matrices $\mathbf{D}_k$ are randomly produced, whose elements are randomly produced with a zero mean unit variance normal distribution for both real and imaginary parts. Furthermore, each column of the mixing matrix $\mathbf{A}$ has been normalized to a unit norm. For making a fair comparison, all the six algorithms are initialized with the identity matrix.

The first experiment investigates the algorithms' robustness to outliers and noise effects. We set $M=N=10$, $K=20$, the set of target matrices contains 20% outliers. Figs. 1-3 show the curves of the convergent GRL, iterative numbers, and CPU time of all the seven algorithms versus NER, respectively. The experimental results demonstrate that after convergence only SDDJD reaches an acceptable performance, which means that the proposed algorithm is robust to outliers. Figs. 4-6 display curves of the performance indexes versus percentage of outliers for $M=N=6$, $K=20$ and $\delta=0.01$. The results indicate that the proposed algorithm achieves a better average performance.

When the percentage of outliers in the target matrices is greater than 60%, the performance of the proposed algorithm degenerates significantly.

In order to reveal more convergence details, we apply the non-orthogonal joint diagonalization under different kind of noise disturbance. Here all the target matrices are given by $\mathbf{R}_k = \mathbf{A}\mathbf{D}_k\mathbf{A}^H + \delta_k \Delta \mathbf{R}_k$, where $M = 6$, $N = 6$, $K = 13$, and $\delta_k = 0.01$ for $k = 1, 2, \ldots, 5$, $\delta_k = 0.02$ for $k = 6, 7, \ldots, 10$, $\delta_k = 1, 2, 3$ for $k = 11, 12, 13$. Fig.7 shows the typical convergence curves of the parameters $\mu_k, k = 1, 2, \ldots, K$ versus the number of iteration. Fig.8 plots the average convergence parameters $\mu_k, k = 1, 2, \ldots, K$ versus the noise level $\delta_k, k = 1, 2, \ldots, K$ over 100 trials. This experiment underlines the usefulness of the weight estimation. The larger the residual, the smaller the weight. Using soft weights, the outliers can be sufficiently detected. That is why the proposed algorithm shows robustness to outliers.

In the second experiment, we compare the different AJD algorithms: PARAFAC [21], BIA [10], CJDi [4], UWEDGE [12], ALS [11] and NOODLES [3] with 2 sources and 2 microphones. The sources have 10-s duration and they are convolved with artificial RIRs generated by the method proposed in [22]. The dimensions of the chosen room are $12m \times 9m \times 3m$. The $x$ and $z$ coordinates of the two sources are fixed to 2 and 1.6, respectively, while the $y$ coordinates are $\{8,1\}$. The $x$ and $z$ coordinates of the two microphones are 10 and 1.6, respectively, while the $y$ coordinates are $\{8,1\}$. $T_p$ is 0.2 s. The FFT length is fixed to 2048 and an overlap coefficient fixed to 75%. Signal-to-interference ratio (SIR) improvement [21] is used as the performance index. The boxplot of 100 experiments is shown in Fig. 9. It can be observed from the simulated experiments that SDDJD has the best performance among the compared approaches.

## 4. Conclusions

In this paper, a novel soft decision-directed non-unitary joint diagonalization algorithm has first been established. The proposed procedure is a valuable tool for improving the performance of separation in the presence of outliers. When the soft weights are employed, outliers can be selectively eliminated, by means that those containing large fitting errors are weighted respectively less during updating. The proposed SDDJD algorithm is derived based on a modified stationary point procedure, whose updating direction is constructed by the linear approximation to the gradient function of the criterion. After establishing this method to estimate the mixing matrix, the diagonal matrices are determined through the standard least-squares optimization procedure. Numerical examples are presented to illustrate that the proposed algorithm possesses potential advantages both in terms of convergent GRL and convergence speed comparing with the state-of-the-art algorithms.

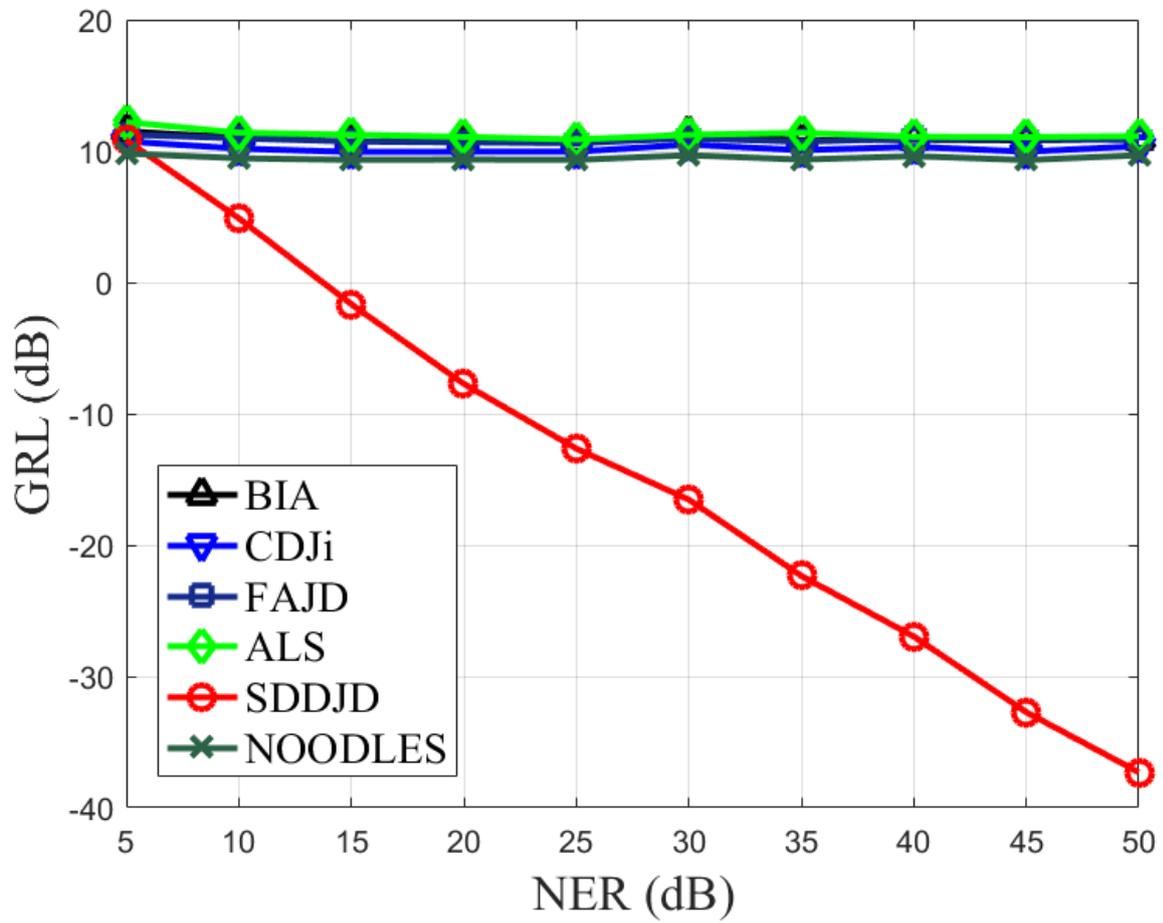

Figure 1: Curves of convergent GRL versus NER.

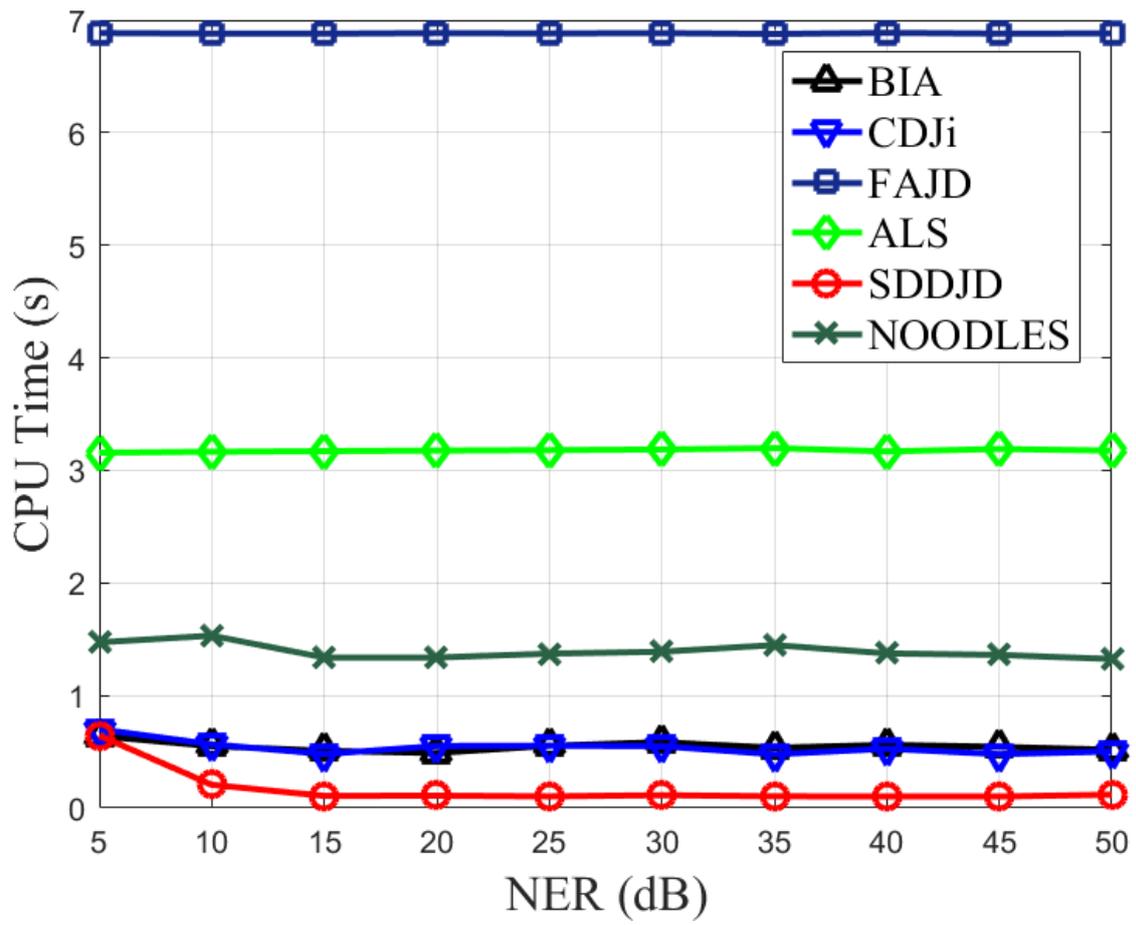

Figure 2: Curves of convergence time versus NER.

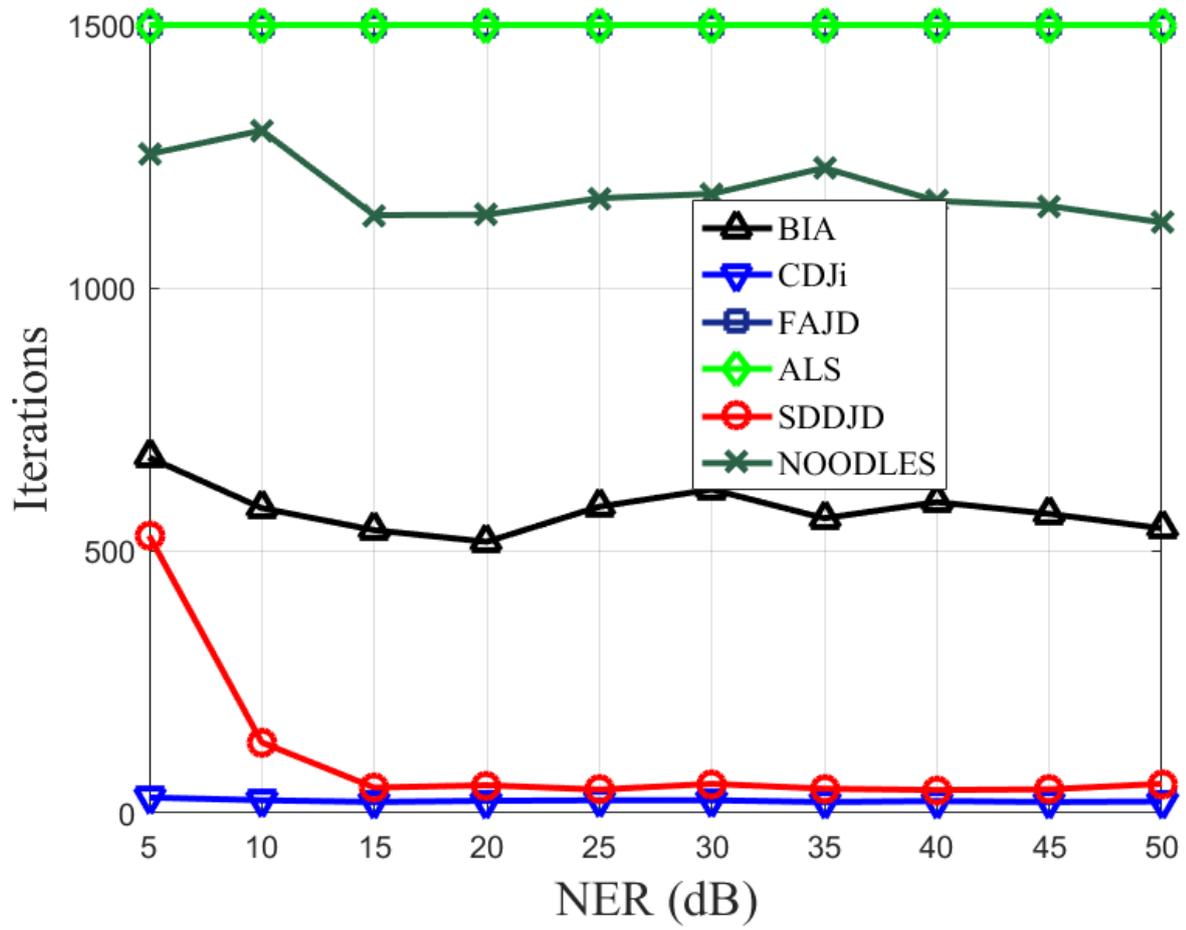

Figure 3: Curves of number of iterations versus NER.

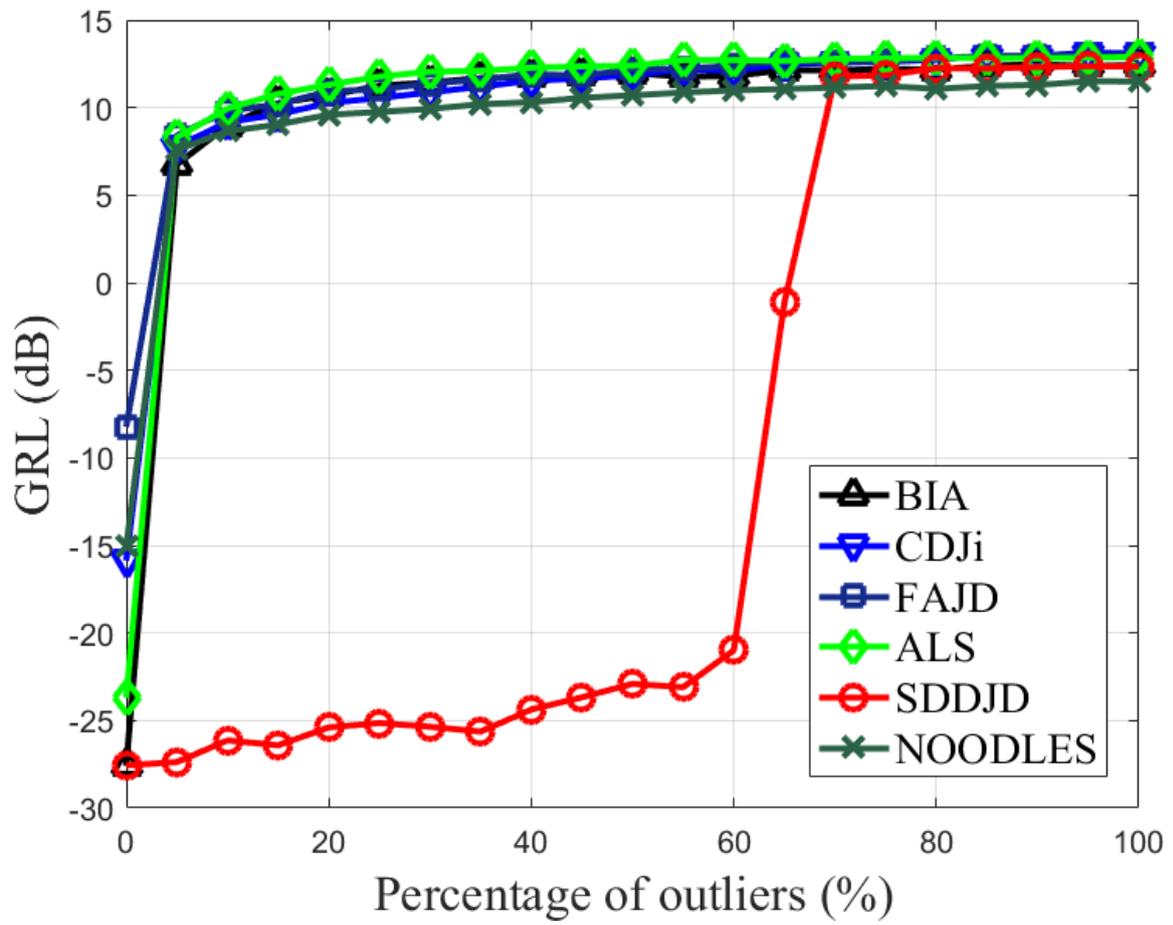

Figure 4: Curves of convergent GRL versus percentage of outliers.

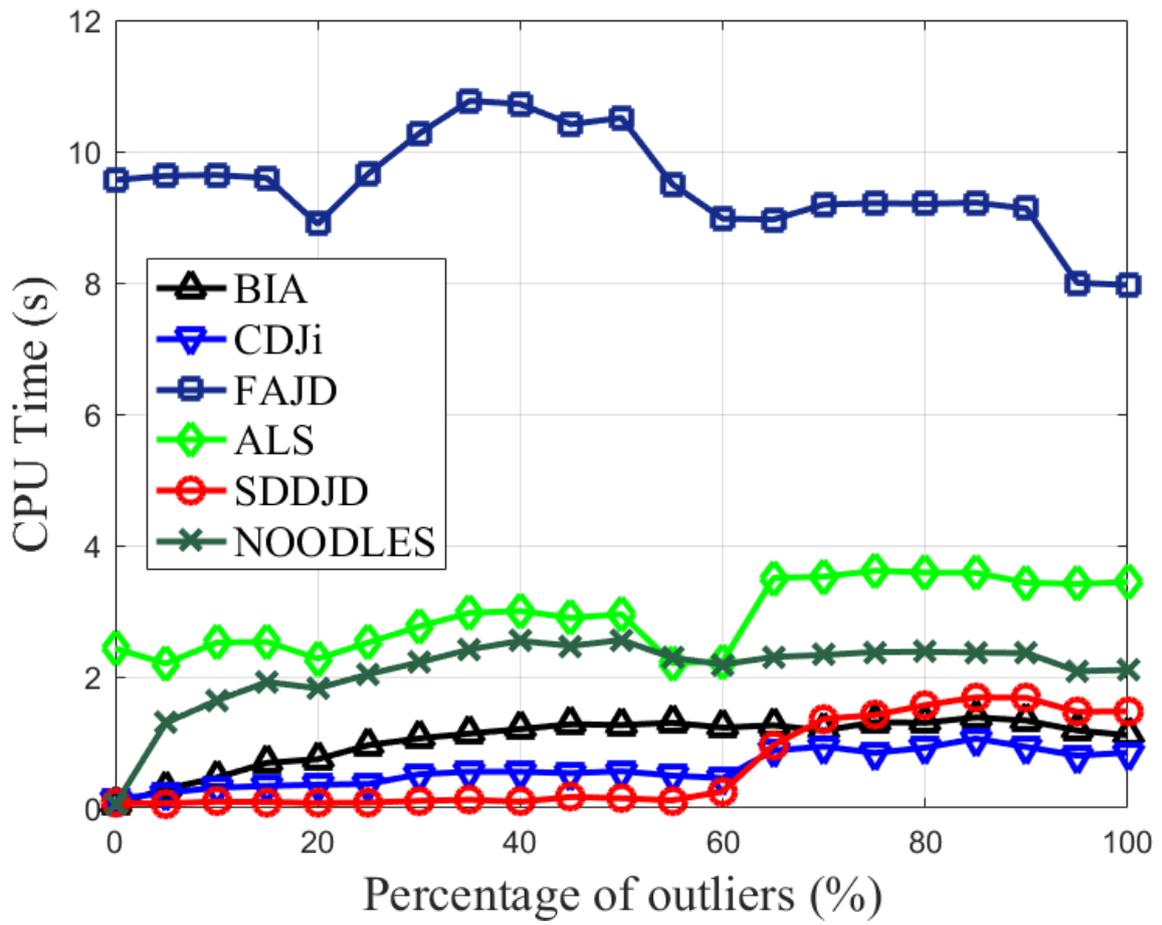

Figure 5: Curves of convergence time versus percentage of outliers.

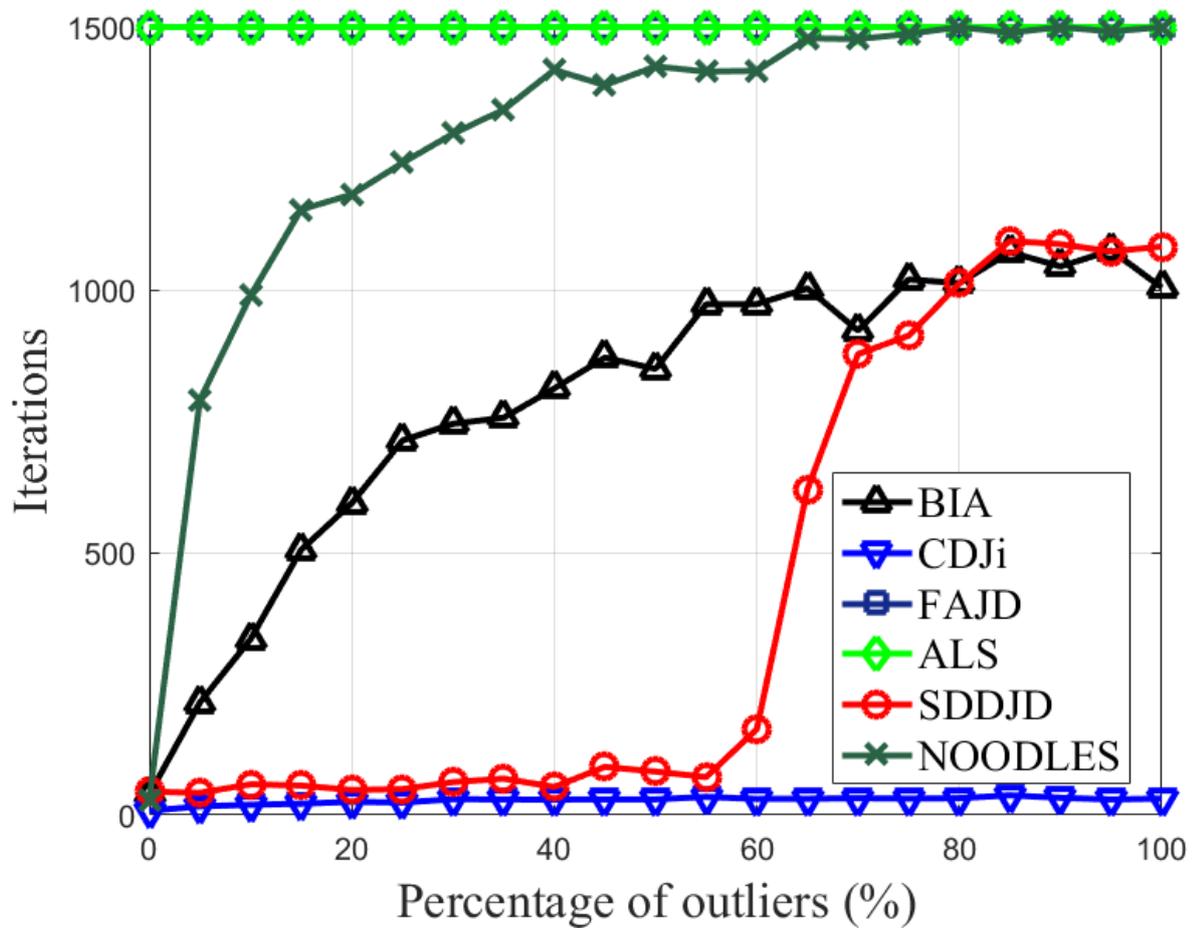

Figure 6: Curves of number of iterations versus percentage of outliers.

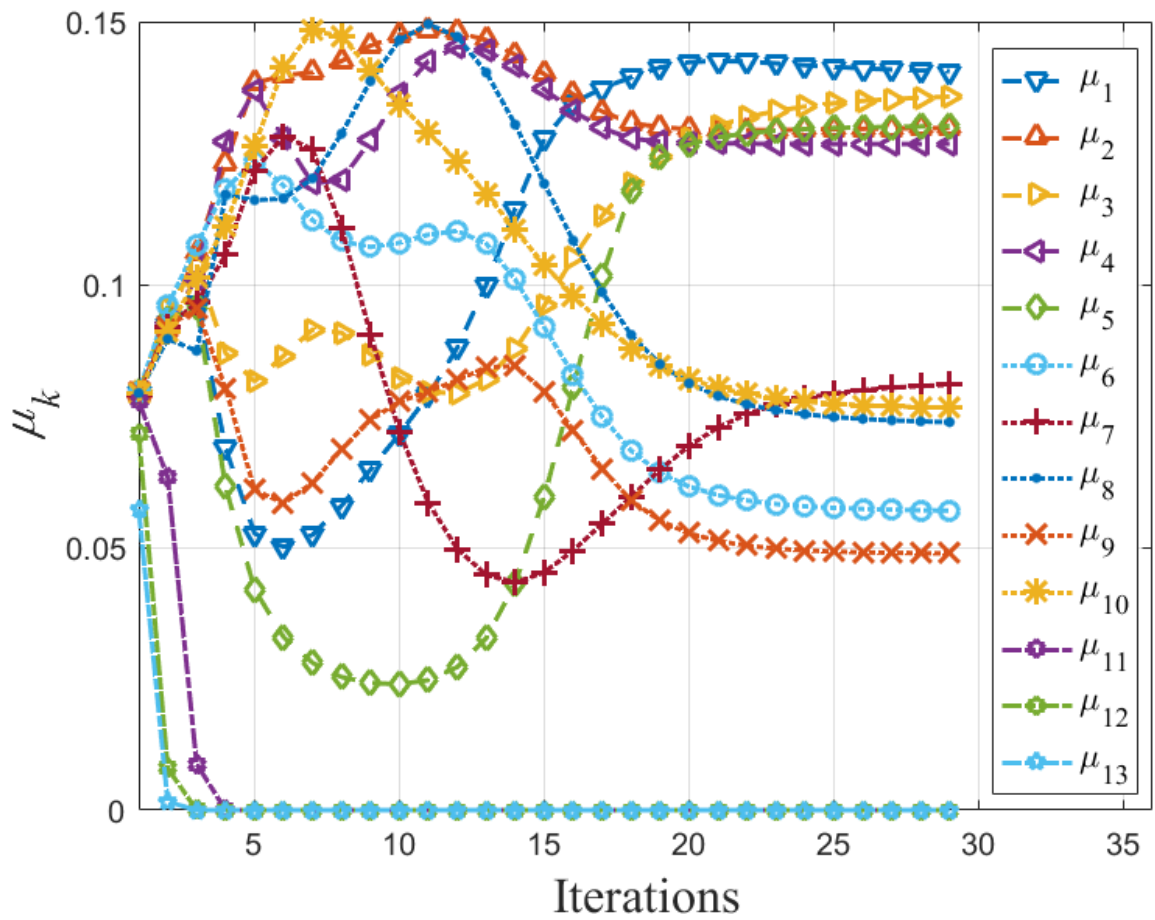

Figure 7: Curves of $\mu_k$ versus iterations.

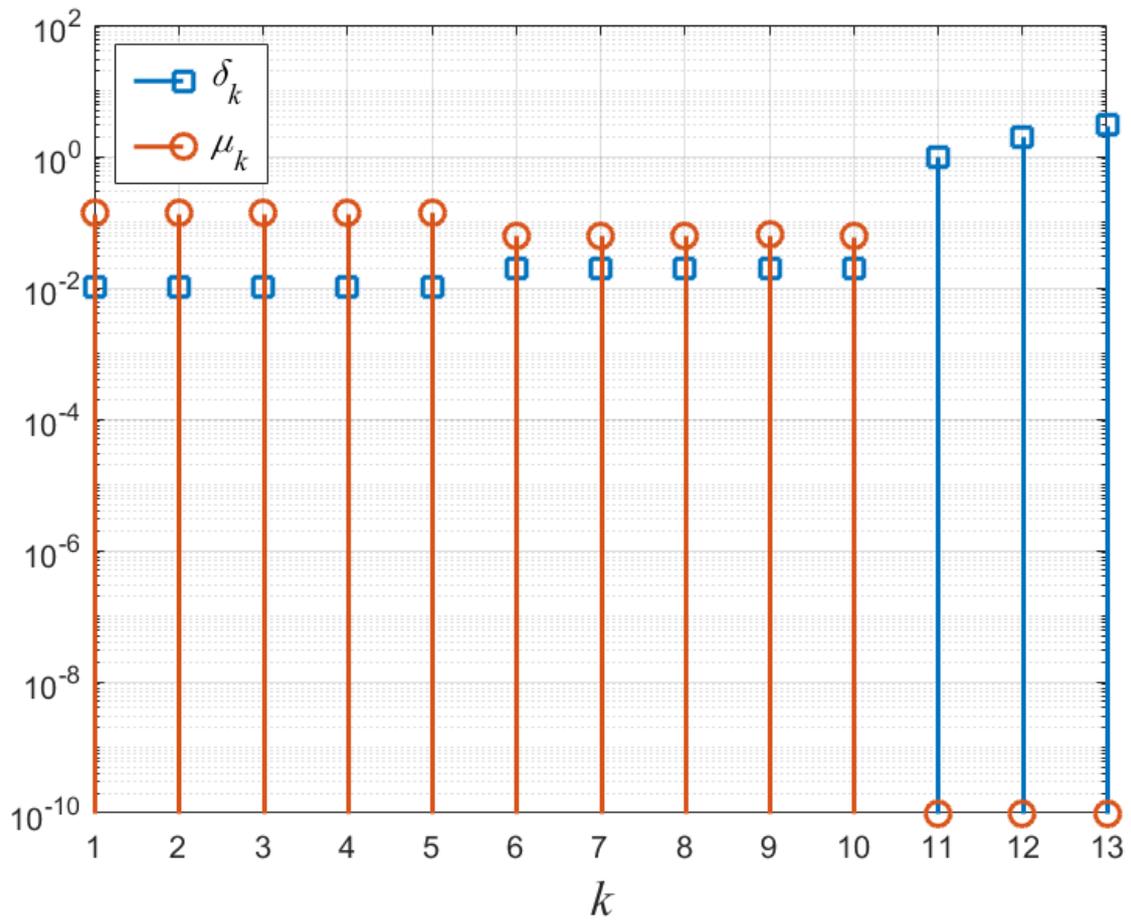

Figure 8: Curves of $\mu_k$ versus $\delta_k$.

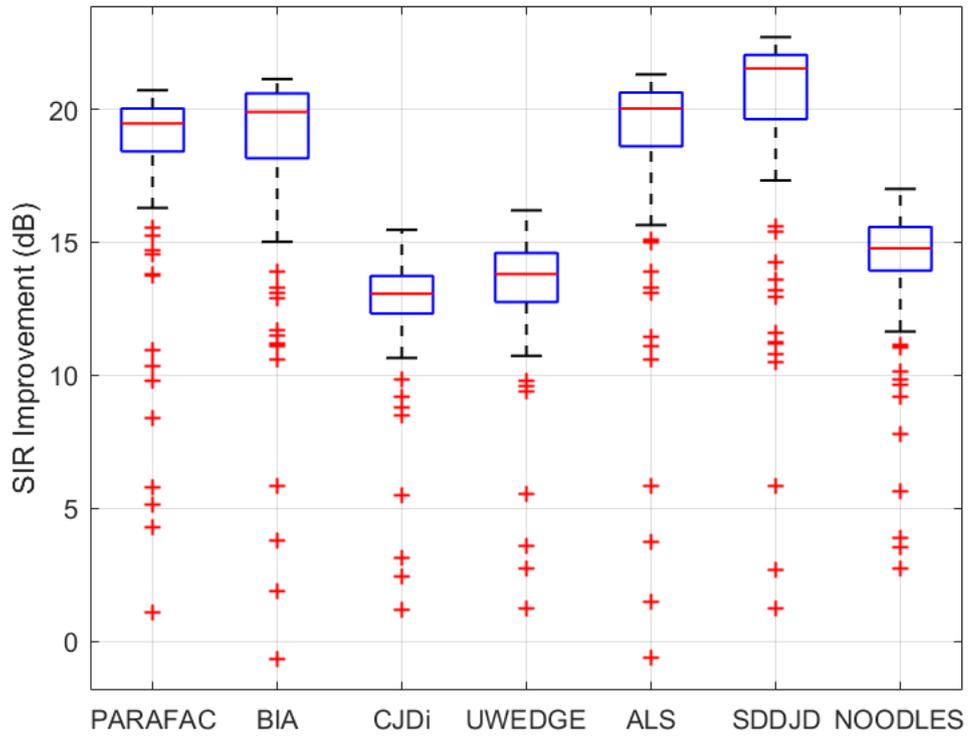

Figure 9: SIR improvement.